\documentclass[onecolumn,preprintnumbers,amsmath,amssymb]{revtex4}
\pdfoutput=1

\usepackage{graphicx}
\usepackage{dcolumn}
\usepackage{bm}

\begin{document}


\title{Second harmonic generation in GaP photonic crystal waveguides}

\author{Kelley Rivoire$^{1*}$, Sonia Buckley$^{1}$, Fariba Hatami$^{2}$, and Jelena Vu\v{c}kovi\'{c}$^1$}

\email{krivoire@stanford.edu} 

\address{$^1$E. L. Ginzton Laboratory, Stanford University, Stanford, CA 94305-4085\\ $^2$Department of Physics, Humboldt University, D-10115, Berlin, Germany}


\begin{abstract}
We demonstrate enhanced second harmonic generation in a gallium phosphide photonic crystal waveguide with a measured external conversion efficiency of 5$\times$10$^{-7}$/W. Our results are promising for frequency conversion of on-chip integrated emitters having broad spectra or large inhomogeneous broadening, as well as for frequency conversion of ultrashort pulses.
\end{abstract}

\maketitle
Nonlinear frequency conversion processes can be strongly enhanced in semiconductor photonic nanocavities and waveguides\cite{corcoran, johnson, bravoabad, burgess, liscidini}, dramatically reducing the required physical footprint and providing an integrated, on-chip platform. Such devices are useful as on-chip light sources at otherwise inaccessible wavelengths, as well as for interfacing between multiple quantum emitters with different frequencies or between quantum emitters and telecommunications wavelengths\cite{mccutcheon_broadband, rivoire_qd}.

Previous work in $\chi^{(2)}$ nanophotonic nonlinear frequency conversion\cite{murray_InP, shg_rivoire_opex, sfg_rivoire} has focused on using nanophotonic cavities with high quality factor \textbf{(Q)} to generate a resonant enhancement in conversion efficiency at the wavelength of the cavity mode; however, the bandwidth allowed by a high-Q cavity would be unsuitable for frequency conversion of quantum emitters with a broadband spectrum (such as the nitrogen vacancy (NV) center) or frequency conversion of ultrashort pulses. Photonic crystal waveguides offer an alternative to cavities; the waveguide photonic dispersion can produce large slow-down factors greatly increasing the effective nonlinear interaction length over a broad wavelength range\cite{corcoran, corcoran2}.
Third harmonic frequency conversion has been previously demonstrated in a silicon photonic crystal waveguide\cite{corcoran}; however, the efficiency is limited by the weak $\chi^{(3)}$ nonlinearity of the material, linear absorption of the third harmonic, and two-photon absorption of the fundamental frequency. Here, we demonstrate second harmonic generation using the $\chi^{(2)}$ nonlinearity of gallium phosphide, which has a large electronic band gap ($\sim$550 nm) minimizing absorption. The circulating intensity in the waveguide is enhanced by the group index $n_g$, providing a factor of $n^2_g$ enhancement in conversion efficiency compared to a conventional waveguide with similar size. This material is compatible with quantum emitters including InP\cite{fariba_qd} and InGaAs quantum dots\cite{yale_qd}, NV centers\cite{dirk_nanoletters}, and fluorescent molecules\cite{molecules}.

Fig. 1 shows the dispersion diagram of transverse electric (TE)-like modes of the W1 waveguide (in-plane E field component only at the center of the slab). The W1 waveguide is formed by removing one row of holes from the triangular lattice\cite{marko_wg}. The photonic band gap is indicated by the white region between the shaded gray areas. The structure supports two guided modes inside the band gap with even (mode of interest) and odd symmetry of the $B_z$ field component relative to the $xz$-plane containing the waveguide axis. The inset shows the field component $B_z$ of the photonic crystal waveguide mode calculated from 3D finite difference time domain (FDTD) simulation at the $k_x$=$\pi/a$ point for the green band (even mirror symmetry). Here, no other changes to the photonic crystal have been made, although group velocity dispersion could be minimized using modified designs shifting the first two rows of holes relative to the defect\cite{krauss}.

Fig. 1b shows a scanning electron microscope (SEM) image of a photonic crystal waveguide, fabricated in a 164 nm thick gallium phosphide membrane. The membrane is grown by gas-source molecular beam epitaxy on top of a 1 $\mu$m thick layer of Al$_{0.85}$Ga$_{0.15}$P on a (100)-oriented gallium phosphide wafer. The structures are fabricated by e-beam lithography and dry etching through the GaP membrane followed by a wet etch to remove the sacrificial layer yielding a suspended membrane. The W1 waveguide is fabricated with lattice constant $a$=560 nm, hole radius $r/a\approx$0.25, and is 30 periods (17 $\mu$m) long. To access wavevectors of the dispersion lying below the light line (which have low loss and large group index) from normal incidence, the structure includes two grating couplers: a circular $\lambda/2n$ grating on the left\cite{andrei_dit} and a coupler on the right formed by perturbing the lattice with periodicity $2a$ using enlarged holes\cite{dirk_incoupler} (used only for transmission measurements). The structures are oriented with the waveguide along a [011] crystal direction. For transmission measurements, an objective lens with numerical aperture of 0.5 is positioned above right grating; the light source is focused slightly off-center onto the left grating. Fig. 2a shows transmission measurements using a tungsten halogen white light source. Periodic peaks in transmission result from the finite length of the waveguide (30 periods) which supports Fabry Perot (FP)-like reflections with peaks separated by $\lambda^2/(2Ln_g)$ where $L$ is the length of the waveguide and $n_g$ is the group index\cite{notomi_wg}. The linewidths of the peaks becomes narrower as the wavelength increases entering the part of the dispersion below the light line. The derived group index is shown in Fig. 2b. The maximum group index measured is 25, lower than expected from simulation (Fig. 2)b, but still much larger than in a conventional waveguide.

To measure second harmonic generation, we couple light from a tunable continuous wave telecommunications-wavelength laser (Agilent 81989A) into the left grating. Second harmonic generated inside the waveguide scatters out of plane and is collected by the same objective placed above the center of the structure from the entire field of view onto a camera for imaging or a monochromator and CCD for spectral analysis. The collected second harmonic radiation as a function of incident laser wavelength is shown in Fig. 2c for a different structure with slightly larger $r/a$ that better matched the range of our tunable laser. Conversion is observed over a range of incident wavelengths, with periodic peaks again observed from FP reflections at the fundamental wavelength. The inset shows the close match between the second harmonic counts at the FP peaks and the expected $n^2_g$ enhancement, where $n_g$ is calculated from the spacing between FP peaks in the second harmonic data. Fig. 2d shows the second harmonic counts measured as a function of incident laser transmitted through the objective. The laser wavelength is 1561.5 nm; the calculated group index for this structure at this wavelength is 30. A log-log fit of the data yields a slope of 1.9, close to what is expected for a second-order process. The estimated external conversion efficiency, calculated from the maximum second harmonic intensity  is 10$^{-9}$ for 2 mW transmitted through the objective, or 5$\times$10$^{-7}$/W. The power at second harmonic wavelength was measured by separately calibrating the spectrometer counts with a laser at the second harmonic wavelength using a power meter. This is four orders of magnitude less than measured for photonic crystal cavities in GaP\cite{shg_rivoire_opex}. As a comparison to previous work on frequency conversion in III-V semiconductors, we note that our device has only ~$70,000$ times lower second harmonic external conversion efficiency for 1.5 $\mu$m light, although it has not been optimized for incoupling or outcoupling efficiency and has a waveguide length 300 times shorter than past experiments (using GaAs/AlGaAs waveguides phase matched by orientation patterning\cite{Yu} that require significantly more complicated MBE growth).

Because the only nonzero elements of the $\chi^{(2)}_{xyz}$ tensor for (100)-grown GaP have $x\neq y \neq z$ and the guided mode of the photonic crystal waveguide at the fundamental wavelength has electric field in-plane only, the second harmonic must have electric field out-of-plane, e.g. couple to transverse magnetic-like (TM-like) mode. A 3D FDTD simulation of electric field at the second harmonic frequency in a periodic triangular photonic crystal lattice is shown in Fig. 3a; the symmetry of the electric fields indicates a monopole mode with far-field radiation pattern (calculated through Fourier transform of the field immediately above the slab\cite{jelena_farfield}) shown in Fig. 3b. The removal of a row of holes to form the W1 photonic crystal waveguide perturbs this mode; the resulting electric field is shown in Fig. 3c, with far field patterns in Figs. 3d-f. The simulated quality factor of the mode is 200, more than an order of magnitude lower than for the same mode unperturbed by the waveguide. To verify experimentally that we collect second harmonic through this mode, we image the radiation pattern from the waveguide onto a camera (Figs. 3g-i) using a Glan Thompson polarizer to resolve different polarizations; the resulting images show very good agreement with the simulations. The wavelength of the incident laser for these measurements is 1561.5 nm ($a/\lambda=0.359$, in good agreement with simulations (Fig. 1)).

In conclusion, we demonstrate enhancement of second harmonic generation in GaP in a photonic crystal waveguide with maximum group index of 30. We measure an external conversion efficiency of 5$\times$10$^{-7}$/W or 10$^{-9}$ for 2 mW incident power. The large electronic band gap of GaP minimizes linear absorption of the second harmonic as well as two-photon absorption of the fundamental. Our results are promising for frequency conversion of on-chip integrated emitters having broad spectra (e.g., NV centers) or large inhomogeneous broadening  (e.g., InAs/GaAs quantum dots), as well as for frequency conversion of ultrashort pulses.

Financial support was provided by the National Science Foundation (NSF Grant ECCS-10 25811 ).  KR and SB supported by Stanford Graduate Fellowships and the NSF GRFP (SB). This work was performed in part at the Stanford Nanofabrication Facility of NNIN supported by the National Science Foundation under Grant No. ECS-9731293.

\begin{figure}[ht]
\includegraphics{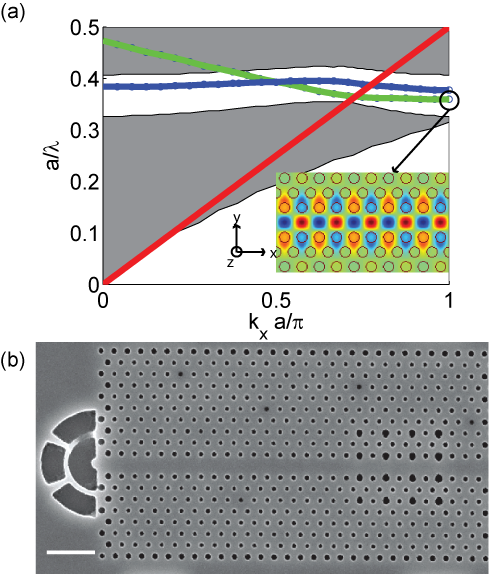}
\caption{\label{fig:figure1}
 (a) Dispersion diagram for TE-like modes (E-field at the center of the slab in plane) for W1 waveguide. Parameters are: hole radius $r/a$=0.25 and thickness $d/a$=0.25 where $a$ is periodicity in $\hat{x}$ direction. White area between gray shaded regions indicates photonic band gap of triangular lattice photonic crystal. Green line indicates waveguide mode with even symmetry of the $B_z$ field component relative to the x-z plane (including waveguide axis). Blue line indicates waveguide mode with odd symmetry. Circles indicate FDTD-calculated solutions and solid lines indicate interpolated bands. Red line indicates light line. The inset shows the FDTD simulation of the $B_z$ field component at the center of the slab for the band plotted in green at the $k_x=\pi/a$ point (circled) with frequency $a/\lambda$=0.36. (b) SEM image of 30-periodic PC photonic crystal waveguide fabricated in 160 nm thick GaP membrane with $a$=560 nm. Scale bar indicates 2$\mu$ m. Circular grating at left or modified holes forming coupler (right) can be used to couple into waveguide modes from free space.}
\end{figure}

\begin{figure}[h]
\includegraphics{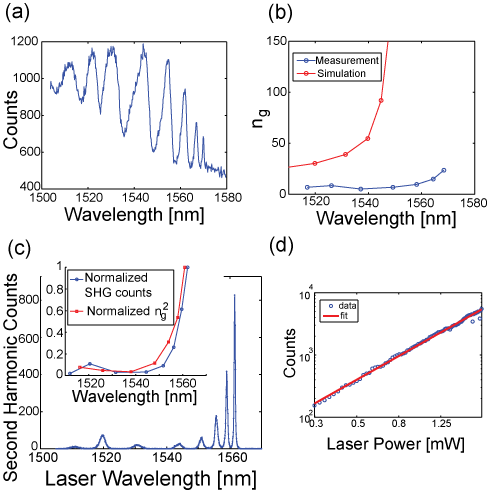}
\caption{\label{fig:figure2}
 (a) Transmission measurement through photonic crystal waveguide using white light source. (b) Group index calculated from FDTD simulation (Fig. 1a) and experimental data (Fig. 2a). Maximum group index measurement experimentally is 25. (c) Second harmonic intensity measured from a different structure as a function of incident laser wavelength. Inset: second harmonic counts at the FP peaks and calculated $n^2_g$ as a function of wavelength. (d) Second harmonic counts measured as a function of incident laser power. Red line indicates linear fit of log-log data with slope 1.9. The laser wavelength is 1561.5 nm; the calculated group index at this wavelength is 30.}
\end{figure}

\begin{figure}[h]
\includegraphics{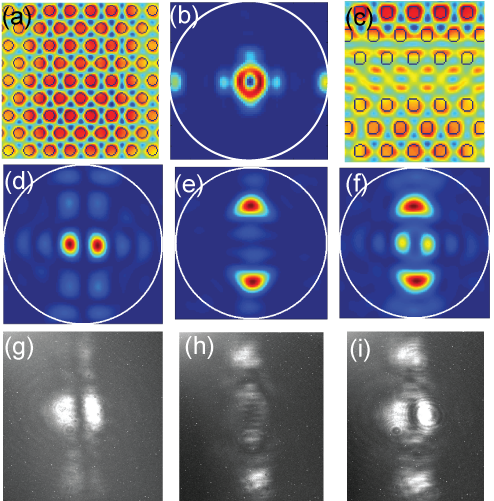}
\caption{\label{fig:figure3}
 (a) FDTD simulation of $E_z$ field component of the monopole TM-like mode near frequency of second harmonic in triangular photonic crystal lattice.  ($a/\lambda$=0.67) (b) Calculated far-field radiation pattern of mode in (a). White circle indicates numerical aperture of lens. (c) FDTD simulation of $E_z$ in photonic crystal waveguide at second harmonic frequency  ($a/\lambda$=0.71). (d) Calculated far-field radiation pattern of $|E_x|^2$ for mode in (c). (e) Calculated far-field radiation pattern of $|E_y|^2$ for mode in (c). (f) Calculated far-field radiation pattern of $|E|^2$ for mode in (c). (g) Measured radiation pattern for $|E_x|^2$. (h) Measured radiation pattern for $|E_y|^2$. (i) Measured radiation pattern $|E|^2$. }
\end{figure}

\end{document}